\def\year{2020}\relax
\documentclass[letterpaper]{article} 
\usepackage{aaai20}  

\usepackage[ruled,linesnumbered ]{algorithm2e}

\usepackage{times}  
\usepackage{helvet} 
\usepackage{courier}  
\usepackage[hyphens]{url}  
\usepackage{graphicx} 
\usepackage{graphicx}  
\usepackage{amsfonts}

\title{Advances in Symmetry Breaking for SAT Modulo Theories}
\author{Saket Dingliwal, Ronak Agarwal, Happy Mittal, Parag Singla\\ 
\textsuperscript{\rm 1}Indian Institute of Technology, Delhi\\ 
\{cs1150254, cs1150252, happy.mittal, parags\}@cse.iitd.ac.in
}

\newtheorem{defn}{Definition}
\newtheorem{thm}{Theorem}
\newtheorem{theorem}{Theorem}

\begin{document}
\maketitle
\begin{abstract}
Symmetry breaking is a popular technique to reduce the search space for SAT solving by exploiting the underlying symmetry over variables and clauses in a formula. The key idea is to first identify sets of assignments which fall in the same symmetry class, and then impose ordering constraints, called Symmetry Breaking Predicates (SBPs), such that only one (or a small subset) of these assignments is allowed to be a solution of the original SAT formula. While this technique has been exploited extensively in the SAT literature, there is little work on using symmetry breaking for SAT Modulo Theories (SMT). In SMT, logical constraints in SAT theories are combined with another set of theory operations defined over non-Boolean variables such as integers, reals etc. SMT solvers typically use a combination of SAT solving techniques augmented with calls to the theory solver. In this work, we take up the advances in SAT symmetry breaking, and apply them to the domain of SMT. Our key technical contribution is the formulation of symmetry breaking over the Boolean skeleton variables, which are placeholders for actual theory operations in SMT solving. These SBPs are then applied over the SAT solving part of the SMT solver. 
We implement our SBP ideas on top of CVC4, which is a state-of-the-art SMT solver. Our approach can result in significantly faster solutions on several benchmark problems compared to the state-of-the-art. Our final solver is a hybrid of the original CVC4 solver, and an SBP based solver, and can solve up to 3.8\% and 3.1\% more problems in the QF\_NIA category of 2018 and 2019 SMT benchmarks, respectively, compared to CVC4, the top performer in this category.



\end{abstract}

\section{Introduction}
\label{sect:introduction}
A large class of AI problems can be reduced to the problem of SAT. One of the important advances in the theory of SAT solving is the notion of symmetry breaking~\cite{crawford&al96}. In symmetry breaking, the key idea is to prune the search space by eliminating possible solutions (and non-solutions) which are symmetrical to each other. Symmetries are defined in terms of (logical) variable permutations which when applied to the theory result in an identity mapping. All the assignments belonging to the same symmetry class are first identified and then an ordering constraint is imposed over them using what is referred to as, Symmetry Breaking Predicates (SBPs). These SBPs result in eliminating a large number of assignments which are identical to each other, often leaving one single assignment in each symmetry class, and therefore drastically pruning the search space. Use of SBPs for SAT solving has been shown to give significant computational advantages for a large class of problems and a number of advances in symmetry breaking have been proposed over time~\cite{aloul&al06,devriendt&bogaerts15}.

Coming from the other direction, there has been a great interest in extending SAT to the class of problems which involve operations like addition, subtraction, comparison etc. over non-Boolean variables (referred to as {\em theory variables}) such as integers, reals, arrays and uninterpreted functions. 
SAT Modulo Theories (SMT) is a framework for constructing decision problems by combining operations over these non-Boolean theory variables along with SAT style logical connectives~\cite{moura&bjorner08}. 
For example, consider a formula: $(x+y > 2) \land (y+z > 2) \Rightarrow (x+y+z >6) \forall x,y,z\in \mathbb{Z}$.
This is an SMT formula defined over integer variables and corresponding operations. SMT theories have a wide range of applications in hardware and software verification, extended static checking, constraint solving, planning, scheduling, test case generation, and computer security~\cite{moura&al07}.
SMT solvers typically work by first creating a {\em Boolean Skeleton} which is obtained by replacing atomic Boolean expressions in the theory by dummy Boolean variables.
A solution to the Boolean skeleton is found and checked for consistency with theory variables. If the solution is found to be inconsistent, then negation of that solution is added to the SMT. The process is repeated until a consistent solution is found. A number of advances in SMT solving have been proposed in the recent literature~\cite{barrett&al18}.

A natural question then arises: can we exploit existing techniques in SAT symmetry breaking to further improve the efficiency of existing state-of-the-art SMT solvers? Unfortunately, the prior work \cite{deharbe&al11} in this direction is limited to the case where the form of symmetries are assumed to be exchangeable symmetries, i.e., when any two variables in a symmetric set can be interchanged with each other without changing the theory. However there may exist other kinds of symmetries which correspond to arbitrary variable permutations. In this work, we further explore the use of SBPs defined over these arbitrary permutations for SMT solving. For instance, in the example considered above, there is no exchangeable symmetry, but we have a symmetry which cyclically rotates the three variables $x,y,z$. Our key insight is to simultaneously find symmetries over the Boolean skeleton of the theory such that these symmetries also correspond to valid symmetries of the theory variables underlying the skeleton. One such symmetry would map the expressions $(x+y > 2)$ to $(y+z >2)$ while at the same time map $x \rightarrow y \rightarrow z \rightarrow x$. We refer to these as {\em Boolean-Skeleton-cum-Theory (BST)} symmetries. We formally define SBPs constructed over BST symmetries, and show that they are sound, i.e., adding them to the theory does not change the satisfiability of the original problem. We also design a number of heuristics to select variable ordering during SBP construction to achieve enhanced performance. Our approach can be seen as a wrapper and can be implemented over any given SMT solver. 

We implement our symmetry based approach over the state-of-the-art CVC4-2018~\cite{barrett&al18} and CVC4-2019~\footnote{\url{https://smt-comp.github.io/2019/system-descriptions/cvc4.pdf}} solvers. These were the top performers in several categories during 2018~\footnote{\url{http://smtcomp.sourceforge.net/2018/results-toc.shtml}} and 2019~\footnote{\url{https://smt-comp.github.io/2019/}} SMT competitions. Our algorithm results in competitive performance compared to the state-of-the-art but does not quite win over them. A careful analysis reveals that the original solver and the symmetry based solver perform better on a fairly different set of problems, leading us to design a simple hybrid solver: we run the original solver for a fraction of time, and if we do not reach the solution, we resort to the symmetry based solver. This results in the best performance with our hybrid approach solving close to $3.8\%$ and $3.1\%$ more problems on 2018 and 2019 benchmarks in the QF\_NIA category of the SMT competition.

\subsection{Contributions}
The main contributions of our work are as follows:
\begin{enumerate}
    \item Our work is the first attempt to exploit a general class of symmetries over the theory variables in SMT problems, as opposed to the earlier approaches which exploit only a subset of symmetries, for example, exchangeable symmetries~\cite{deharbe&al11}.
    \item We propose the novel idea of exploiting SAT style symmetries over Boolean skeleton variables in tandem with symmetries over the theory variables in SMT problems.
    \item We propose an ensemble/hybrid solver, which combines the power of existing non-symmetry based solvers, with our symmetry based solver.
    \item We show that our hybrid solver beats the state-of-the-art solver on several benchmark problems solving close to $3.8\%$ and $3.1\%$ more problems in the QF\_NIA category of SMT benchmarks from 2018 and 2019 competitions, respectively.
\end{enumerate}
The outline of this paper is as follows. First, we give some background on Symmetry breaking and SMT solving. We then review the existing literature on symmetry breaking, and advances in SMT solving. We then describe our framework for introducing SBPs in SMT including the proof of soundness. We then propose variable ordering heuristics to improve the efficiency of our method. This is followed by our experimental results on various benchmark problems. We conclude with the directions for future work.

\section{Background}
\label{sect:background}
\subsection{Symmetry Breaking}
Consider a SAT theory $\mathcal{T}$ defined over a set of Boolean variables $\mathcal{X}=\{X_1,X_2,X_3\cdots X_n\}$. Let $\theta:\mathcal{X}\to\mathcal{X}$ denote a permutation over the variables in the set $\mathcal{X}$. Typically, permutations are denoted using the cyclic notation. For example, $(X_1 X_2 \cdots X_k) (X_{k+1} X_{k+2} \cdots X_{l}) \ldots (X_j,X_{j+1},\cdots,X_n)$ denotes that $X_1$ maps to $X_2$, $X_2$ to $X_3$, and so on, until $X_k$ which maps to $X_1$. This denotes one cycle in the permutation. Similar mapping holds for each cycle, i.e., variables enclosed between round brackets. These mappings should hold simultaneously.
\begin{defn}
A permutation $\theta$ is a symmetry of theory $\mathcal{T}$ if application of $\theta$ over $\mathcal{T}$ results back in $\mathcal{T}$. 
\end{defn}
For example, consider a SAT theory: $a\lor b, a \lor c, b \lor d$. Then the permutation $\theta=(a b)(c d)$ is a symmetry of this theory.
It is easy to see that the set $\Theta$ of all symmetries of a theory form a group, i.e., they have an identity, each element has an inverse, and they are closed under composition. 

Consider an assignment $\mathcal{A}:\mathcal{X}\to 2^n$. Then the application of permutation $\theta$ on $\mathcal{A}$, is defined as $\theta(\mathcal{A}(v))=\mathcal{A}(\theta(v))$, where $v\in\mathcal{X}$. Given an assignment $x$, its orbit is defined as $\{x' | \exists \theta, \theta(x) =x'\}$ where $\theta$ is a symmetry.

As shown in \cite{crawford&al96}, symmetry detection for SAT problems can be converted to an equivalent graph-isomorphism problem. SAUCY 3.0~\cite{katebi&al12} is a common available tool used for finding the symmetry permutation given a SAT problem. 

The goal of symmetry breaking is to introduce constraints such that only one of the assignments in any orbit is allowed to be a solution. Typically, the chosen assignment is the lexicographically smallest assignment in its orbit (under some variable ordering). This is done by adding a predicate of the following form to the theory:
\begin{equation}
 \wedge_{1 \le i \le n} (\wedge_{1 \le j < i} X_j = \theta(X_j)) \Rightarrow \left(X_i \Rightarrow \theta(X_i)\right)
\end{equation}
These are known as Symmetry Breaking Predicates (SBPs)~\cite{crawford&al96}. \\

\subsection{Satisfiability Modulo Theories (SMT)} 
SMT problems are extension of SAT problems to allow operations over non-Boolean types such as integers, reals, bit-vectors, and arrays etc. An SMT formula is constructed by defining operations over what are referred to as {\em theory variables}, and combining them with logical connectives. Each variable takes values from its domain, and operations over the variables include standard operations such as addition and subtraction over integers, and also include comparison, equality and inequality. The last three operations result in Boolean expressions which can be combined using logical connectives. Given an SMT formula $\Omega$ it is said to be satisfiable if $\exists$ an assignment to theory variables which satisfies the formula. For example, the SMT formula in the introduction is satisfiable under the assignment $x=3,y=3,z=3$. 

In the theory of SMT solving, any given formula $\Omega$ can be thought of composed of two kinds of constraints (a) Logical constraints (denoted by $\Psi$) over atomic Boolean expressions appearing in the formula (b) Consistency constraints (denoted by $\Phi$) capturing the definition of each atomic Boolean expression. Then, $\Omega=\Psi \wedge \Phi$. $\Psi$ is referred to as the {\em Boolean skeleton} of the original formula $\Omega$. This way of decomposing the constraints forms the basis of modern day SMT solvers. First, a solution $x$ to the Boolean skeleton is found using a SAT solver, which is then checked for consistency in $\Phi$ using a theory solver. If a consistent solution $u$ to the theory variables is found, it is returned and formula is declared SAT. If found inconsistent, a conflict clause $C$ (forcing negation of $x$) is added to $\Omega$ and process is repeated until a consistent solution is found. When no further repetition can be made, the formula is declared UNSAT. Algorithm 1 provides a high-level understanding of the SMT solving relevant to our technique. Most of the modern day SMT solvers Yices~\cite{dutertre14}, VeriT~\cite{bouton&al09}, CVC4~\cite{barrett&al11} use the same basic algorithm.

\begin{algorithm}[]
\SetKwInput{KwData}{Input}
\SetAlgoLined
\KwData{SMT $\Omega$}
\KwResult{Model $M$ for $\Omega$ or UNSAT} 
 $\Psi \leftarrow $ Boolean-Skeleton ($\Omega$);\\
 $\Phi \leftarrow$ Constraint-Set($\Omega$);\\
 $x \leftarrow $ SAT-Solve ($\Psi$);\\
 \While{$\Psi$ is SAT}{
  \eIf{T-consistent($\Phi, x$)}{
   $u \leftarrow$ T-Model($\Phi$)\;
   return $u$\;
   }{
    $C \leftarrow$ Conflict-Clause($\Phi, x$);\\
    $\Psi \leftarrow \Psi \land C$ ;\\
    $x \leftarrow $ SAT-Solve ($\Psi$);\\
  }
 }
 return UNSAT\;
 \caption{SMT solving algorithm}
\end{algorithm}

\section{Related work}
\label{sect:related}
There is a vast existing literature on exploiting the symmetries for SAT problems. Building on the original work by Crawford~\shortcite{crawford&al96}, there have been several attempts for using symmetry breaking to improve the the efficiency of SAT solvers. Torlak \& Jackson~\cite{torlak&jackson07} adopted the approach of first finding the symmetries, and then adding the corresponding symmetry breaking clauses in the theory to reduce the search space. Sakallah~\shortcite{sakallah09} proposed symmetry breaking techniques for propositional logic, which improved the performance of SAT solvers. The symmetries are generally detected by first converting the problem to graph automorphism problem, and then using the efficient graph automorphism detection tools like Saucy~\cite{katebi&al12} and Nauty~\cite{mckay&piperno14}. In the space of Constraint Satisfaction Problems (CSPs), Cohen et. al.~\shortcite{cohen&al05} identified two kinds of symmetries that can be exploited: (a) Solution symmetries, preserving the solutions of the problem, (b) Constraint symmeteries, preserving the constraints of the problem. Backofen \& Will~\shortcite{backofen&will99} proposed symmetry breaking techniques for CSPs. However, there has been limited work in the direction of exploiting symmetries in SMT problems. Deharbe et al.~\cite{deharbe&al11} exploit a limited class of symmetries, known as exchangeable symmetries over theory variables, i.e., where all the variables in a symmetric set of completely interchangeable with each other. SyMT~\cite{areces&al13} was developed as a tool to find symmetries over variables in an SMT problem. It is based on reduction of the problem to graph automorphism which is then solved by using existing solvers such as Saucy~\cite{katebi&al12}. However, to the best of our knowledge, none of the existing techniques have exploited symmetry breaking in the SAT component of the SMT solver. 
In this paper, we present first such technique, which exploits symmetries breaking in the SAT module of the SMT solvers, resulting in significant gains over the state-of-the-art. There is also some existing work~\cite{chaganty&al13} which exploits symmetries in conjunction with an SMT solver for inference in Markov logic networks~\cite{richardson&domingos06}, but they do not directly work on exploiting symmetries within an existing SMT problem, and their technique is limited to use of lifting~\cite{kimmig&al15}, which is different from our usage of SBPs.
\section{Exploiting Symmetries in SMT}
\label{sect:theory}
Let us first describe a formulation for constructing SBPs for a given SMT formula $\Omega$.

\subsection{Preliminaries}
Let the variables in SMT be given by the set $\{Y_1,Y_2\cdots,Y_m\}$. Note that these variables could come from different types. We assume that given a variable type $t$, the values come from a domain $D_t=\{v_1,v_2,\cdots v_k\}$. We allow both finite domains as well as countably infinite domains, with the assumption that there is a natural (total) ordering between the values in the domain, e.g., for integer type, this is the natural ordering between integers. For finite valued domains, this could be arbitrarily imposed. 

We will also assume that the variable set in the SMT is an ordered set, and without loss of generality, we will assume that there is some pre-defined ordering over types, and all the variables of a certain type $t_i$ appear before all the variables of another type $t_j$ in the variable ordering if type $t_i$ appears before type $t_j$ in the type ordering. 

\subsection{Constructing SBPs: Basics}
Next, we construct a colored graph automorphism problem for variables and constraints over them in $\Omega$ using the algorithm described in~\cite{areces&al13} (see Background Section). Let $\theta$ denote the symmetry. Then, an SBP can be constructed as follows:
\begin{equation}
 \wedge_{1 \le i \le m} (\wedge_{1 \le j < i} Y_j = \theta(Y_j)) \Rightarrow \left(Y_i \Rightarrow \theta(Y_i)\right)
\end{equation}
This is identical to the SBP described in the Background section with the difference that now we are dealing with non-Boolean variables. In the above equation, we say that $Y_i \Rightarrow Y_k$ iff the value $v_i$ taken by the variable $Y_i$ and the value $v_k$ taken by the variable $Y_k$ (they should have identical domains) are such that $v_i \le v_k$. This ordering is well defined due to the assumption about ordering among the domain values of any given type. Correctness follows from the correctness of SBPs in the Boolean case.

\begin{thm}
Let $\Omega$ be an SMT theory. Let $\theta$ be a symmetry over it. Consider the theory $\Omega'=\Omega \wedge SBP(\theta)$. $\Omega$ is satisfiable iff $\Omega'$ is satisfiable.
\end{thm} 
The proof follows from the proof of correctness of SBPs in the logical case; since there is nothing in the proof which assumes operations over logical variables only.

\subsection{Exploiting Boolean Skeleton Variables}
{\bf Intuition:}
Before we go to formal description of our approach for adding SBPs in an SMT formulation, let us get some intuition. Existing literature on SBPs already tells us how to exploit symmetries over purely logical formulas. Further, the tool SyMT extends the symmetries over Boolean formulas to the variables in an SMT formula (see section on related work). Then, one straight forward approach would be to define the SBP over symmetries discovered over these SMT formulas as described above. A similar approach has been taken by existing work~\cite{deharbe&al11}, though they focus only on exchangeable variable symmetries (i.e., when all the variables in a symmetric set can be freely exchanged with each other). In addition to working only with a restricted symmetry class, this approach has another limitation: it has no way of exploiting symmetries over the Boolean skeleton variables. Since in any given SMT solver, a significant chunk of computation is taken by the SAT solver over the Boolean skeleton, the symmetries discovered in~\cite{deharbe&al11} does not help in any improvement over this component. 

We take a slightly different approach to exploit symmetry breaking in SMT. Instead of discovering symmetries only for theory variables, we instead construct a graph automorphism problem to simultaneously discover symmetries over the Boolean skeleton variables as well as the theory variables. Since these symmetries are discovered jointly, this means that any such symmetric permutation can be applied to get a new configuration of the Boolean skeleton variables as well as the theory variables. Then, we exploit the symmetry over the Boolean skeleton variables to construct an SBP over the corresponding logical formula. These symmetries can then be incorporated in the SAT component of the SMT solver. Next, we describe our approach in detail.

\noindent
{\bf Algorithm:}
Consider an SMT theory $\Omega$. Let $\Omega$ be decomposed as $\Omega=\Psi \wedge \Phi$, where $\Psi$ and $\Phi$ denote the Boolean skeleton and theory constraints, respectively. Then, clearly, checking satisfiability over the original theory $\Omega$ is equivalent to checking satisfiability over the RHS above. Since we would like to work with SBPs formed over Boolean skeleton variables, we define the notion of a restricted SBP as follows:
\begin{defn}
{\bf Restricted SBPs:} Let $t$ be a variable type in an SMT formula. Let $\theta$ be a symmetry of SMT. We construct the restricted SBP for type $t$ variables as follows: we order the types such that $t$ appears first in the ordering. Correspondingly, the variables for type $t$ appear first in the variable ordering. Let these variables be given by  $\{Y_1,Y_2,\cdots, Y_r\}$. Then, the restricted SBP is defined as:
\begin{equation}
 \wedge_{1 \le i \le r} (\wedge_{1 \le j < i} Y_j = \theta(Y_j)) \Rightarrow Y_i \Rightarrow \theta(Y_i)
\end{equation}
\end{defn}
The only difference from the standard SBP formulation (Equation 2) is that now the ordering constraints over only variables from the type $t$ are considered. Clearly, adding restricted SBPs to a theory still preserves satisfiability. Given this definition, we construct an SBP using the following steps:
\begin{enumerate}
\item Construct a colored graph automormphism problem corresponding to the SMT problem expressed by $\Psi \wedge \Phi$. Let $\Theta$ denote the set of symmetries discovered by colored graph automorphism. Initially, we assign $\Psi' \leftarrow \Psi$.
\item For every symmetry $\theta \in \Theta$, construct the restricted SBP corresponding to Boolean skeleton variables. Add this SBP to the SMT theory: $\Psi' \wedge$ SBP-form($\theta$) $\wedge \Phi$. Note we add the newly constructd SBP specifically to the Boolean skeleton part since these SBPs are defined only over the variables in the Boolean skeleton by construction.
\item Let the final SMT theory after adding SBPs be given as $\Psi' \wedge \Phi$. Then, we now apply our SMT solver over the this newly developed theory. Note that SBPs have been added over the Boolean skeleton variables which will be exploited during the SAT computation phase of SMT. Consistency constraints have remained unchanged.
\end{enumerate}
Algorithm 2 gives the pseudocode for the entire SMT solving using our SBP based approach.
\begin{theorem}
The SMT problem represented by $(\Psi' \land \Phi)$ and $(\Psi \land \Phi)$ are equi-satisfiable i.e., if one is satisfiable then other is also satisfiable, and vice-versa. Here, the notations are as used in Algorithm 2.
\end{theorem}

We note that it is a design choice to enforce SBPs only over the Boolean skeleton variables. We primarily motivated by the fact that most of the search (DPLL style) happens here, and therefore, adding SBPs to this phase is quite important in pruning the search space. This is also demonstrated by the gain obtained in our experimental results. We can of course, extend our SBPs to also be constructed over remaining theory variables, experimenting with them is a direction for future work.

\begin{algorithm} 
\SetAlgoLined
\SetKwInput{KwData}{Input}
\KwData{SMT $\Omega$}
\KwResult{Model $M$ for $\Omega$ or UNSAT} 
 $\Psi \leftarrow $ Boolean-Skeleton ($\Omega$);\\
 $\Phi \leftarrow$ Constraint-Set($\Omega$);\\
 $\theta_1,\theta_2\cdots\theta_n \leftarrow$ Symmetry-Permutations($\Psi \land \Phi$);\\
 $\Psi' \leftarrow \Psi$; \\
 \For{$i\gets0$ \KwTo $n$ }{
    $\Psi' \leftarrow \Psi' \land$ SBP-Form($\theta_i$) ;\\   
  }
 $x \leftarrow $ SAT-Solve ($\Psi$);\\
 \While{$\Psi'$ is SAT}{
  \eIf{T-consistent($\Phi, x$)}{
   $u \leftarrow$ T-Model($\Phi$)\;
   return $u$\;
   }{
    $C \leftarrow$ Conflict-Clause($\Phi, x$);\\
    $\Psi \leftarrow \Psi \land C$ ;\\
    $x \leftarrow $ SAT-Solve ($\Psi$);\\
  }
 return UNSAT; \\
} 
 \caption{SMT solving algorithm + Symmetry Breaking Technique}
\label{alg:smt-sym}

\end{algorithm}

\noindent
{\bf Example:}
Table 1 provides an example with a  problem with linear integer theory. The SBP ($\mathcal{T}$) in this example is $Q\Rightarrow R$, under the variable ordering $P,Q,R,S,T$. The second 
assignment is selected over the first one (with first one being pruned after SBP addition)  
and hence the solver do not have to explore both of them. 
\begin{table*}[t]
\centering
\caption{Symmetry Breaking Technique}\smallskip
\begin{tabular}{ll}
SMT Problem  $\Omega$                       & $((z > 2) \lor (x < 8)) \land ((z > 2) \lor (y < 8)) \land ((x+y < 10) \lor (x+y > 3)) $                                               \\
Boolean Skeleton  $\Psi$                    & $(P \lor Q) \land (P \lor R) \land (S \lor T)$                                                                           \\
Constraints Set  $\Phi$                     & $(P \Leftrightarrow (z > 2)) \land (Q \Leftrightarrow (x < 8)) \land (R \Leftrightarrow (y < 8)) \land (S \Leftrightarrow (x + y < 10)) \land (T \Leftrightarrow (x + y > 3))$ \\
Satisfying Assignment &  $z=3; x=7; y=9; P=True; Q=True; R=False; S=False; T=True$ \\
Symmetry Permut.  $\theta$                 & $\theta(Q) = R$;  $\theta(R) = Q$; $\theta(x) = y; \theta(y) = x$                                                   \\
Satisfying Assignment  & $ z=3; x=9; y=7; P=True; Q=False; R=True; S=False; T=True $\\
(after applying $\theta$) & \\
SBP added                               & $Q \Rightarrow R$                                                                                       \\
New Skeleton $\Psi$ $'$ & $(P \lor Q) \land (P \lor R) \land (S \lor T) \land (\neg Q \lor R)$ \\
Assignment chosen & $ z=3; x=9; y=7; P=True; Q=False; R=True; S=False; T=True $\\
(lexicographically smaller one) & \\
\end{tabular}
\label{table1}
\end{table*}

In order to maximize the impact of SBPs formed and reduce overheads we introduce various heuristics explained in the subsections below.  

\subsection{Variable Ordering Heuristic}
The symmetry breaking technique is invariant of any arbitrary variable ordering chosen. Also, once the ordering is chosen, it must remain consistent for all the symmetries detected. However, to maximize the effect of the process, we tried different heuristics on the variable ordering. The key idea here is to make the new clause added to play a significant role in deciding the truth assignment for the Boolean skeleton. We choose the variables appearing as positive unit variables in the clausal normal form of the Boolean skeleton to appear first in the variable ordering. The main intuition behind such a choice lies the fact that most SAT solvers first assign true values to these unit variables for unit propagation, and hence for the Symmetry Breaking Predicate of the form $Y_i \Rightarrow \theta(Y_i)$ , the value of $\theta(Y_i)$ gets automatically decided. The SAT solver will now directly assign $True$ value to the $\theta(Y_i)$ and hence pruning the branch where it could have been assigned a $False$. This simple heuristic boosts the performance of the technique discussed here.

\subsection{Reduced Size Heuristic}
The step mentioned in line 6 of the Algorithm 2 involve forming SBP from the found symmetry permutation. Although we mention the general form of SBP added in the theory section, but we are free to choose any subset of that SBP without breaking the soundness. As suggested by~\cite{aloul&al03} in the Symmetry Breaking for SAT solving, breaking all symmetries may not speed up the search because there are often exponentially many of them and may be redundant. We restrict to single symmetry generators given by graph-automorphism solver of~\cite{areces&al13}. Further, it was seen that smaller generator cycles seems to be more effective. Therefore, we restrict the size of permutations larger than k bits to k, where k is a hyperparameter.   
\subsection{Hybrid Technique}
In our experiments, we found that there are a lot of non-overlapping problems solved by algorithm 1 and Algorithm 2 in the SMT benchmarks. Particularly, for the 2018 QF\_NIA benchmarks, we saw that algorithm 2 is able to solve large number of benchmarks that algorithm 1 was not able to solve. However, there was another equally large chunk of problems in which algorithm 1 beat algorithm 2. Hence, the overall gain in the number of benchmarks solved was quite low. This means that for several problems adding SBPs resulted in significant speed-ups, whereas in other cases, it slowed down the original solver. We hypothesize this is because in the former case, unnecessary parts of the search space are pruned resulting in speed-up, whereas in the latter case, the SBPs might be pruning valid models, resulting in further delay in reaching a satisfying assignment. 

Taking insight from this, we design a hybrid solver, where we first run the solver with SBPs added for a fixed time $t$, and if the solution is still not found, we resort to the original solver with SBPs removed. $t$ is a hyperparameter. 
This results in significant boost in performance compared to CVC4 on a subset of SMT benchmarks. 
\section{Experiments}
\label{sect:experiments}
\subsection{Goal}
Our experiments are designed to help us answer two important questions about our algorithm. The first one is whether our algorithm, when implemented on top of an existing SMT solver, is able to improve its performance. And the second one is to check whether our solver can push the state-of-the-art in solving benchmark SMT problems. Both of these goals are essential to show that our algorithm is effective as well as independent of the solver. 

\subsection{Benchmarks}
We use both 2018~\footnote{\url{http://smtcomp.sourceforge.net/2018/benchmarks.shtml}} and 2019~\footnote{\url{https://smt-comp.github.io/2019/benchmarks.html}} version of the benchmarks of the SMT-COMP. These benchmarks are categorized based on the underlying theory. We use the category QF\_NIA particularly because the theory is relatively complex and as per our hypothesis, shows some promising results. These benchmarks have a mix of SAT and UNSAT problems. 

\subsection{Solvers}
We implemented all the symmetry breaking techniques, hybrid technique and heuristics discussed in this paper to build our solver CVC4-SymBreak. 
CVC4-SymBreak-18 and CVC4-SymBreak-19 are two versions of our solver built on CVC4-18~\cite{barrett&al18} and CVC4-19~\footnote{\url{https://smt-comp.github.io/2019/system-descriptions/cvc4.pdf}} respectively. We participated as a non-competing entry in SMT-COMP- 2019~\cite{barth&al19} with our solver CVC4-SymBreak-18. SMT-COMP is an annual competition of all the SMT solvers on a set of SMT problems called benchmarks which are updated every year. Although, the technique is invariant of the underlying solver, but we choose CVC4-18~\cite{barrett&al18} here as it was winner in many categories in the SMT-COMP-2018 and is publicly available. For detecting the symmetries of the problem, we use the tool SyMT~\cite{areces&al13} and then added the corresponding predicates(SBPs) formed in the Boolean skeleton of the problem generated by CVC4. We compare our performance particularly with CVC4 to achieve our first goal. Also, participation in the competition helps us to know our position vs the sate-of-the-art. 

\subsection{Methodology}
We ran CVC4-18, CVC4-19 and our solvers on various benchmarks in different categories and sub-categories. All the experiments were conducted in the same setting as used by the SMT-COMP for the competition. Some of them are direct results from the SMT-COMP-19 while others were run separately. The wall-clock timeout was 1200 seconds and 2400 seconds for each individual problem in 2018 and 2019 benchmarks respectively. The memory was limited to 60 GB in both the experiments. All the experiments were run on the starexec cluster~\cite{stump&al14} with Intel(R) Xeon(R) CPU E5-2667 v4 @ 3.20GHz and CentOS Linux release 7.4.1708.

\subsection{Results}
The results are tabulated in tables 2 and 3 for 2018 and 2019 versions of benchmarks and solvers respectively. The aim of the solver is to solve a problem in a given time constraint (the wall-clock timeout). As mentioned in the previous subsection, the time limits for each benchmark were set as per the rules of SMT-COMP. Therefore, metric used to evaluate performance of a solver is the number of benchmarks solved. Our solver with symmetry breaking was able to solve 905 ($3.8\%$ of the total benchmarks) more problems than its counterpart in SAT-COMP-2018. Similarly, on SMT-COMP-2019 benchmarks, our solver was able to solve 549 and 358 ($4.8\%$ and $3.1\%$ of the total benchmarks) more problems than its counterparts. Hence we see that we are able to answer positively to both the questions which we talked above. 
This shows that our technique is versatile and can build on improved performance of the underlying solvers. 
In other words, we believe that ideas proposed here are fundamental in nature, and have the capacity to benefit even new solvers that might be proposed in future. 
Further, both CVC4-2018 and CVC4-2019 are the state-of-the-art for the benchmarks of their respective year. 
Although, the winner of the latest SMT-COMP, solver Par4~\footnote{\url{https://smt-comp.github.io/2019/participants/par4}} is able to solve more problems than us in this category but it achieves so by using different solvers in parallel for solving the problem. In the sequential setting, our results are the new state-of-the-art.  




\begin{table}[t]
\centering
\begin{tabular}{|l|r|}
\hline
\textbf{Solver Name} & \textbf{\#Problems}\\
\hline
CVC4-18 & 15662 \\
CVC4-SymBreak-18 & 16567 \\
\hline
\textbf{Total Bechmarks} & 23876\\
\hline
\end{tabular}
\caption{Number of solved problems on 2018 benchmarks}
\label{table2}
\end{table}

\begin{table}[t]
\centering
\begin{tabular}{|l|r|r|r|}
\hline
\textbf{Solver Name} & 
\multicolumn{3}{c|}{
\textbf{\#Problems}}  \\
\hline
& SAT & UNSAT & ALL\\
\hline
CVC4-18 & 5101 & 2286 & 7387 \\
CVC4-SymBreak-18 & 5542 & 2394 & 7936 \\
CVC4-19 & 5785 & 2504 & 8289 \\
CVC4-SymBreak-19 & 6093 & 2554 & 8647 \\
\hline
\textbf{Total Benchmarks} & & & 11494 \\
\hline

\end{tabular}
\caption{Number of solved problems on 2019 benchmarks}
\label{table3}
\end{table}

\section{Conclusion}
In this paper, we have presented an approach for incorporating symmetry breaking predicates (SBPs) in SMT problems. Our approach is based on jointly exploiting the symmetries over the Boolean skeleton as well as the theory consistency formulas in the SMT formulation. Our novelty lies in introducing the symmetry breaking predicates for Boolean skeleton variables of SMT, thus improving the SAT module of SMT solvers, which typically is a bottleneck. We prove that our SBP formulation is sound. Our experiments against the state of the art solvers show significant gains on the benchmarks from SMT Competitions of 2018 and 2019. Directions for future work include experimenting with additional benchmarks, and trying our approach with other solvers.

\section{Acknowledgements}
\label{sect:acknowledgements}
We are thankful to Kuldeep Meel and Supratik Chakraborty for helpful discussions and inputs on the work. Happy Mittal was supported by the TCS Research Scholar Program. Parag Singla is supported by the DARPA Explainable Artificial Intelligence (XAI) Program \#N66001-17-2-4032, IBM SUR award, and Visvesvaraya Young Faculty Fellowships by Govt. of India. Any opinions, findings, conclusions or recommendations expressed in this paper are those of the authors and do not necessarily reflect the views or official policies, either expressed or implied, of the funding agencies.
\bibliographystyle{aaai}
\bibliography{all}
\end{document}